\documentclass[fleqn,twoside]{article}
\usepackage{espcrc2}

\usepackage{graphicx}
\usepackage[figuresright]{rotating}

\newcommand{\AmS}{{\protect\the\textfont2
  A\kern-.1667em\lower.5ex\hbox{M}\kern-.125emS}}

\title{The spin structure function of the neutron}

\author{Anthony~W.~Thomas\addressmark\thanks{University of Adelaide
preprint: ADP-01-43/T475; \hspace{1cm}
Invited presentation at the Workshop on the
Spin Structure of the Proton
and Polarized Collider Physics, ECT* Trento, July 23-28, 2001}
\address[CSSM]{Special Research Centre for the Subatomic Structure of
Matter \\ and Department of Physics and Mathematical Physics, \\ 
        The University of Adelaide, Adelaide SA 5005, Australia}}
\begin{document}

\begin{abstract}
The neutron spin structure function, $g_{1n}$, has been of considerable
interest recently in connection with the Bjorken sum rule and the proton
spin crisis. Work on this problem has concentrated on measurements at
low-$x$. We recall the important, non-perturbative
physics to be learnt by going instead to larger
values of $x$ and especially from a determination of the place where the
expected sign change occurs. Of course, in order to obtain neutron data
one must use nuclear targets and apply appropriate corrections. In this
regard, we review recent progress concerning the various nuclear
corrections that must be applied to measurements on polarised $^3$He.
\vspace{1pc}
\end{abstract}

\maketitle

\section{INTRODUCTION}
Since the original discovery of the 
``proton spin crisis'' \cite{Ashman:1988hv} there has
been enormous progress in our knowledge and understanding of the spin
structure functions of the nucleon \cite{Thomas:2001kw}.
It is well known that through the U(1) axial anomaly polarised gluons
can contribute to the spin structure function of the 
proton \cite{Altarelli:1988nr,reviews}  
and that this is quite likely the source of most of the discepancy with
the naive Ellis-Jaffe sum rule. Tests of this idea are now focussed on
direct measurements of the polarised gluon content at Hermes, COMPASS
and RHIC. 

In comparison with the Ellis-Jaffe sum rule, the Bjorken sum rule is 
still on a very sound theoretical foundation and provides a crucial test
of our understanding of QCD. In order to check it one needs data for the
neutron spin structure function, a task complicated by the absence of
free neutron targets. Traditionally neutron data has been extracted from
data on the deuteron \cite{SMC,E143,E155} 
by subtracting the proton contribution. Apart from
Fermi motion and binding corrections this is complicated by the presence
of completely new terms not present for a free nucleon 
\cite{Melnitchouk:1995tx} -- although numerical estimates suggest that these
terms are negligible for a system as weakly bound as the deuteron.
The main problem with the deuteron is that the $n$ and $p$ spins are
aligned so that one has to subtract two numbers of comparable size 
($g_{1p}$ from $g_{1D}$) to get the neutron data.

An attractive alternative is to use 
polarised $^3$He \cite{HERMES,E154}, where the spins of
the two protons are mainly coupled to zero, so that the neutron carries
most of the nuclear spin. As a consequence, the corrections to be applied to
the data to obtain the neutron structure function are expected to be
much smaller than for the deuteron. Yet, binding and Fermi motion are
larger, as are shadowing corrections and meson exchange currents --
especially those involving the $\Delta$. The shadowing corrections
are especially
significant in the small-$x$ region, where most of the contribution 
from polarised gluons is also concentrated. In section 3 we
outline recent progress
in the calculation of those corrections which are relevant at
intermediate and large-$x$. First, however, we
recall the main non-perturbative aspects of nucleon structure that can
be tested by extending measurements to larger values of $x$ than have so
far been accessible. 

\section{LARGE-$X$ AND THE NON-PERTURBATIVE STRUCTURE OF THE NUCLEON}
While the emphasis for the last decade has been on extending our
knowledge of spin structure functions to smaller $x$, we stress that
there is a lot to be learnt by heading in the opposite direction.
Large-$x$ has proven to be a source of surprises for 
unpolarised structure functions, with recent 
analysis \cite{Melnitchouk:1996fc} suggesting that
the $d/u$ ratio may not go to zero (unlike the standard parametrisations
of parton distributions universally assume) but is rather 
consistent with the predictions of perturbative 
QCD (pQCD) \cite{Farrar:1975yb}. Further experiments are planned at
places like JLab to test this analysis \cite{he3_test}.

In the spin dependent case we have no idea whatsoever of the behaviour
of $g_{1n}$ beyond $x \sim 0.4$. On the basis of both pQCD {\it and}
SU(6) \cite{Close:1988br}, one expects $A_{1n}$ to
approach 1 as $x \rightarrow 1$ \cite{Melnitchouk:1996zg,Isgur:1999yb}. 
It is vital
to test this prediction. If it fails we understand nothing about the
valence spin structure of the nucleon! Looking in a little more detail,
we see that the present data at small-$x$ corresponds to a negative
asymmetry and hence there must be a crossover at some intermediate $x$
value. Locating the crossover is an important experimental 
challenge. From the theoretical point of view the value of $x$ at which
this occurs is the result of a competition between the SU(6) valence
structure \cite{Close:1988br} and the chiral 
corrections \cite{Schreiber:1988uw,Steffens:1995at}.

The positive sign of the neutron 
asymmetry in the larger $x$ region is a result of the dominance 
there of $S=0$ spectator pairs
(which have lower mass than $S=1$ pairs) \cite{Close:1988br}. Within the
SU(6) framework, the only valence quark with a spin-0 pair of spectators
is a $d$-quark which has its spin aligned with the spin of the neutron
-- and hence the asymmetry is positive. This 
phenomenon has been studied at great length, not only for the spin-flavor
dependence of nucleon parton distributions \cite{Londergan:1998ai} but
also for the parton distributions and fragmentation functions of other
members of the nucleon octet \cite{Alberg:1996hq,Boros}. The competition
comes from the $N \pi$ and $\Delta \pi$ Fock components
of the wave function of the nucleon \cite{Steffens:1995at},
corresponding to the leading and  next-to-leading non-analytic chiral
behaviour of the structure function \cite{Thomas:2000ny}. In view of the
interest in di-quark models of the nucleon \cite{Alkofer:2001ne}, 
as well as the role of dynamical symmetry breaking, it is imperative to
have as much experimental guidance as possible and the insight from
accurate data on $g_{1n}$ at intermediate and  large $x$ would be
extremely valuable.

\section{EXTRACTING $G_{1n}$ FROM $^3$HE DATA}
Within the usual impulse approximation, 
$g_{1}^{^3{\rm He}}$ can be  represented as the convolution of the 
neutron ($g_{1}^{n}$) and proton ($g_{1}^{p}$) spin structure functions 
with the spin-dependent nucleon light-cone momentum 
distributions $\Delta f_{N/^3{\rm He}}(y)$, 
where $y$ is the ratio of the struck nucleon to nucleus light-cone plus 
components of the momenta 
\begin{eqnarray}
&&g_{1}^{^3{\rm He}}(x,Q^2)=
\int_{x}^{3} \frac{dy}{y} \Delta f_{n/^3{\rm He}}(y)g_{1}^{n}(x/y,Q^2)
\nonumber \\ && +
\int_{x}^{3} \frac{dy}{y} \Delta f_{p/^3{\rm He}}(y)g_{1}^{p}(x/y,Q^2) \ .
\label{conv1}
\end{eqnarray}
The motion of the nucleons inside the nucleus (Fermi motion) and 
their binding  are parametrized through the  
distributions $\Delta f_{N/^3{\rm He}}$, which
can be readily calculated using the ground-state wave function 
of $^3$He. Detailed calculations \cite{CSPS93,SS93,Bissey01}
by various groups. using different ground-state wave functions for $^3$He,  
have come to similar conclusions, namely that $\Delta f_{N/^3{\rm He}}(y)$ 
is sharply peaked around $y \approx 1$ 
because of the small average separation energy per nucleon. 
Thus, Eq.~(\ref{conv1}) is often approximated by
\begin{equation}
g_{1}^{^3{\rm He}}(x,Q^2)=P_{n} g_{1}^{n}(x,Q^2)+ 2P_{p} g_{1}^{p}(x,Q^2) \ ,
\label{conv2}
\end{equation}
where $P_{n}$ ($P_{p}$) are the effective polarizations of 
the neutron (proton) inside polarized $^3$He, defined by
\begin{equation}
P_{n,p}=\int_{0}^{3} dy \Delta f_{n,p/^3{\rm He}}(y) \ .
\label{pnp}
\end{equation}

In the first approximation to the ground-state wave function of $^3$He, 
only the neutron is polarized. 
In this case, $P_n$=1 and $P_p$=0. 
Realistic approaches to the wave function of $^3$He  include also  
higher partial waves, notably the $D$ and $S^{\prime}$ partial waves. 
This leads to the depolarization of the spin of the neutron and polarization 
of protons in $^3$He. The average of calculations 
with several models of the nucleon-nucleon interaction 
and three-nucleon forces can be summarized
as  $P_{n}=0.86 \pm 0.02$ and $P_{p}=-0.028 \pm 0.004$ \cite{FGPBC}. 
The calculations of \cite{Bissey01} give similar values: 
$P_{n}=0.879$ and $P_{p}=-0.021$ for the PEST potential with 5 channels. 
We shall use  these values for $P_{n}$ and $P_{p}$ throughout this paper. 
Most of the uncertainty in the 
values for $P_{n}$ and $P_{p}$ comes from the uncertainty 
in the $D$-wave of the $^3$He wave function. Thus, for the observables 
that are especially sensitive to the poorly constrained $P_{p}$, 
any theoretical predictions carry a significant uncertainty. 
One example of such an observable is the point where the neutron 
structure function $g_{1}^{n}$ has a node.
 
\section{ROLE OF THE $\Delta(1232)$}

The description of the nucleus as a mere
collection of protons and neutrons is incomplete. In polarized DIS 
on the tri-nucleon system, this observation  can be  illustrated  
by the following example \cite{FGS96}.
The Bjorken sum rule relates the difference 
of the  first moments of the proton and neutron spin structure functions 
to the axial vector coupling constant of the neutron,   
$g_{A}=1.2670 \pm 0.0035$ \cite{Thomas:2001kw},
\begin{equation}
\int_{0}^{1}\Big(g_{1}^{p}(x,Q^2)-g_{1}^{n}(x,Q^2)\Big)dx 
=\frac{1}{6}g_{A}\Big(1+O(\frac{\alpha_{s}}{\pi})\Big) \ .
\label{delta1}
\end{equation}
Here the QCD radiative corrections are denoted as ``$O(\alpha_{s} / \pi)$''. 
This sum rule can be straightforwardly generalized 
to the $^3$He-$^3$H system,
with $g_{A}|_{triton}$, the axial vector coupling constant of the 
triton ($g_{A}|_{triton}=1.211 \pm 0.002$ \cite{Budick})
replacing $g_{A}$.
Taking the ratio of the Bjorken sum rules for $A=3$ and the nucleon,
one obtains
\begin{eqnarray}
\frac{\int_{0}^{3}\Big(g_{1}^{^3{\rm H}}(x,Q^2)-g_{1}^{^3{\rm He}}(x,Q^2)\Big)
dx}{\int_{0}^{1}\Big(g_{1}^{p}(x,Q^2)-g_{1}^{n}(x,Q^2)\Big)dx}
&=& \nonumber \\
\frac{g_{A}|_{triton}}{g_{A}}=0.956 \pm 0.004 \ .
\label{delta3}
\end{eqnarray}
Note that the QCD radiative corrections cancel exactly in Eq.\ (\ref{delta3}).

Assuming charge symmetry between the $^3$He and $^3$H ground-state 
wave functions, one can write the triton spin structure function 
$g_{1}(x,Q^2)$ in the form:
\begin{eqnarray}
&&g_{1}^{^3{\rm H}}(x,Q^2)=\int_{x}^{3} \frac{dy}{y} \Delta f_{n/^3{\rm He}}(y)
\tilde{g}_{1}^{p}(x/y,Q^2) \nonumber \\
&+&\int_{x}^{3} \frac{dy}{y} 
\Delta f_{p/^3{\rm He}}(y)\tilde{g}_{1}^{n}(x/y,Q^2) \ .
\label{delta4}
\end{eqnarray}
This leads to the following estimate for the ratio of the nuclear 
to nucleon Bjorken sum rules
\begin{eqnarray}
&&\frac{\int_{0}^{3}\Big(g_{1}^{^3{\rm H}}(x,Q^2)-g_{1}^{^3{\rm He}}(x,Q^2)
\Big)dx}{\int_{0}^{1}\Big(g_{1}^{p}(x,Q^2)-g_{1}^{n}(x,Q^2)\Big)dx}
\nonumber \\ &=&\Big(P_n-2P_p\Big) \frac{\tilde{\Gamma}_{p}-\tilde{\Gamma}_{n}}{\Gamma_{p}-\Gamma_{n}}=0.921 \frac{\tilde{\Gamma}_{p}-\tilde{\Gamma}_{n}}{\Gamma_{p}-\Gamma_{n}}   \ .
\label{delta5}
\end{eqnarray}
Here we used $P_{n}=0.879$ and $P_{p}=-0.021$, and   
$\tilde{\Gamma}_{N}=\int^{1}_{0} dx \tilde{g}_{1}^{N}(x)$ 
and $\Gamma_{N}=\int^{1}_{0} dx g_{1}^{N}(x)$ are, respectively, the
spin sums for bound and free nucleons. 
Such off-shell corrections were estimated within the framework of the
quark-meson coupling model \cite{QMC} in Ref.~\cite{Steffens:1995at}. 
They were not large and showed a tendency to decrease,
rather than increase, the bound nucleon spin structure functions 
(i.e. $(\tilde{\Gamma}_{p}-\tilde{\Gamma}_{n})/(\Gamma_{p}-\Gamma_{n}) < 1$). 
Thus, one can immediately see that the theoretical prediction for the ratio 
of the Bjorken sum rule for the $A=3$ and $A=1$ systems 
(Eq.~(\ref{delta5})), based solely on nucleonic degrees of freedom, 
underestimates the experimental result for the same ratio 
(Eq.~(\ref{delta3})) by 
about 3.5\%.  This  demonstrates the need for new nuclear effects.

It has been known for a long time that non-nucleonic degrees of freedom, 
such as pions, vector mesons, the $\Delta$(1232) isobar, 
can play an important role in the calculation of some low-energy observables in  
nuclear physics. In particular, the analysis of Ref.~\cite{Saito} 
demonstrated that 
the two-body exchange currents involving a $\Delta$--isobar 
increase the theoretical prediction for the axial vector coupling 
constant of the triton by about 4\%,
which makes it consistent with experiment. In order to preserve 
the Bjorken sum rule, 
exactly the same mechanism must be present in case of deep inelastic 
scattering on polarized $^3$He and $^3$H. Indeed, as explained in 
Refs.~\cite{FGS96,BGST},
the direct correspondence between the calculations of the 
Gamow-Teller matrix element in the triton $\beta$ decay and the Feynman 
diagrams for DIS on $^3$He and $^3$H (see Fig.~1 of \cite{BGST}) 
requires that two-body exchange currents should play an equal role in 
both  processes. As a result, the presence of the $\Delta$ in the $^3$He 
and $^3$H wave functions should increase the ratio of Eq.~(\ref{delta5}) and 
make it consistent with Eq.~(\ref{delta3}). 

The contribution of the $\Delta$(1232) to $g_{1}^{^3{\rm He}}$ is realized 
through diagrams involving the non-diagonal interference 
transitions $n \rightarrow \Delta^{0}$ and 
$p \rightarrow \Delta^{+}$. This requires new spin structure 
{}functions, $g_{1}^{n \rightarrow \Delta^{0}}$ and 
$g_{1}^{p \rightarrow \Delta^{+}}$, as well as the effective 
polarizations $P_{n \rightarrow \Delta^{0}}$ and 
$P_{p \rightarrow \Delta^{+}}$. 
Taking into account the interference transitions leads to a correction
to the $A=3$ spin structure functions, $\delta g_1$: 
\begin{eqnarray}
&&\delta g_{1}= \nonumber \\
&\pm& \left[ 2P_{n \rightarrow \Delta^{0}}g_{1}^{n \rightarrow \Delta^{0}}+
4P_{p \rightarrow \Delta^{+}}g_{1}^{p \rightarrow \Delta^{+}} \right],
\label{delta6}
\end{eqnarray} 
where the $\pm$ signs correspond to $^3$He/H, respectively.
 
The interference structure functions can be related to $g_{1}^{p}$ and 
$g_{1}^{n}$ within the quark parton model using the general structure of the 
SU(6) wave functions \cite{Close:1988br,Boros}
\begin{equation}
g_{1}^{n \rightarrow \Delta^{0}}=
g_{1}^{p \rightarrow \Delta^{+}}=
\frac{2 \sqrt{2}}{5}\Big(g_{1}^{p}-4 g_{1}^{n} \Big) \ .
\label{su6}
\end{equation}
This simple relationship is valid in the range of $x$ and $Q^2$ where the 
contribution of sea quarks and gluons to $g_{1}^{N}$ can be safely omitted, 
i.e. for $0.5 \leq Q^2 \leq 5$ GeV$^2$ and $0.2 \leq x \leq 0.8$ if the 
parametrization of Ref.~\cite{GRSV} is used. 

In principle, the effective polarizations of the interference 
contributions $P_{n \rightarrow \Delta^{0}}$ and 
$P_{p \rightarrow \Delta^{+}}$ can be calculated using a $^3$He 
wave function that includes the $\Delta$ resonance. This is an involved 
computational problem. Instead, we chose to find 
$P_{n \rightarrow \Delta^{0}}$ and $P_{p \rightarrow \Delta^{+}}$ 
by {\it requiring} that the
use of the $^3$He and $^3$H structure functions of Eq.~(\ref{delta6}) 
gives the experimental ratio of the nuclear to nucleon Bjorken  
sum rules (\ref{delta3}). This yields
the effective polarizations \cite{Bissey:2001cw}:
\begin{equation}
2\Big(P_{n \rightarrow \Delta^{0}}+2P_{p \rightarrow \Delta^{+}}\Big)=-0.025 \ .
\label{effpol}
\end{equation}
(Note that this value is very close to that 
reported in Ref.~\cite{BGST}.)

\begin{figure}[t]
\begin{center}
\includegraphics[height=16pc]{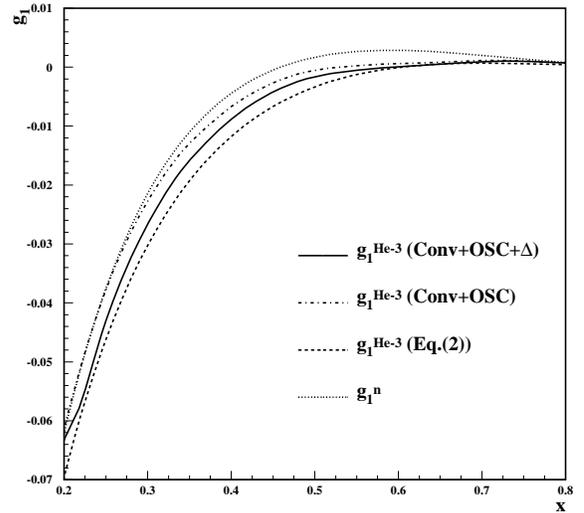}
\caption{The full spin structure function, $g_{1}^{^3{\rm He}}$ (solid curve), 
compared with that computed by allowing for Fermi motion and binding,
with (dash-dotted curve) and without (dashed curve) the inclusion of
off-shell corrections (OSC), estimated in the QMC model.
The free neutron spin 
structure function, $g_{1}^{n}$, is shown by the dotted 
curve. For all curves $Q^2$=4 GeV$^2$.}
\label{fig:delta}
\end{center}
\end{figure}

Clearly one can now write an 
explicit expression for the $^3$He spin structure function, which 
takes into account the additional diagrams corresponding to the 
non-diagonal interference $n \rightarrow \Delta^{0}$ and  
$p \rightarrow \Delta^{+}$ transitions;  
thus ensuring agreement with the experimental value of the ratio of the 
Bjorken sum rules.
The results of such a calculation of $g_{1}^{^3{\rm He}}$ at 
$Q^2=4$ GeV$^2$ are presented in Fig.~\ref{fig:delta} as the solid curve.
One can see from Fig.~\ref{fig:delta} that the presence of the $\Delta$(1232) 
isobar in the $^3$He wave function works to decrease $g_{1}^{^3{\rm He}}$ 
relative to the prediction of Fermi motion and binding alone. This decrease is 
12\% at $x=0.2$ and increases at larger $x$, peaking for $x \approx 0.46$, 
where $g_{1}^{n}$ changes sign.

We note that, since the convolution formalism 
implies {\it incoherent} scattering off 
nucleons and nucleon resonances
of the target, coherent nuclear effects present at small values of 
Bjorken $x$ are being ignored here. In Ref.~\cite{BGST}, the role  
played by two coherent effects, namely nuclear shadowing and antishadowing, 
in DIS on polarized $^3$He were also considered. 
As these corrections are not
significant in the large-$x$ region, which is of concern to us here, 
we pursue this question no further.

\subsection{Correction to the neutron asymmetry at large-$x$} 
The DIS asymmetry $A_{1}^{T}$ for any target $T$ is proportional to the 
spin structure function $g_{1}^{T}$:
\begin{equation}
g_{1}^{T}=\frac{F_{2}^{T}}{2 x(1+R)} A_{1}^{T} \ ,
\label{a1}
\end{equation}
where $R=(F_{2}^{T}-2x F_{1}^{T})/(2xF_{1}^{T})$ and  
$F_{1,2}^{T}$ are inclusive spin-averaged structure functions.   
It is assumed in Eq.~(\ref{a1}) that the transverse spin asymmetry, 
$A_{2}^{n}$, is negligibly small and that $R$ does not depend on the 
choice of target.

Applying this definition of $A_{1}^{T}$ to the $^3$He, proton and neutron 
targets and including the effects of Fermi motion, binding and
$\Delta$--isobars outlined above, 
one obtains for the neutron asymmetry $A_{1}^{n}$
\begin{eqnarray}
&&A_{1}^{n}=\frac{F_{2}^{^3{\rm He}}}{P_{n} F_{2}^{n} 
(1+\frac{0.056}{P_{n}})} \nonumber \\
&&\Bigg(A_{1}^{^3{\rm He}}-2
\frac{F_{2}^{p}}{F_{2}^{^3{\rm He}}}P_{p}A_{1}^{p} 
\bigg(1-\frac{0.014}{2 P_{p}}\bigg)\Bigg) \ .
\label{a1n}
\end{eqnarray} 
Provided that the proton asymmetry, $A_{1}^{p}$, is constrained 
well by the experimental data, the largest theoretical uncertainty 
comes from the uncertainty 
in the proton spin polarization $P_{p}$. We estimate that the 
uncertainty in the second term of  Eq.~(\ref{a1n}) and, thus, in the 
position of the point where $A_{1}^{n}$ has a zero, 
is of the order 10\% \cite{Bissey:2001cw}. 

In Eq.(\ref{a1n}) the terms proportional to $0.056$ and $0.014$ represent the correction to  
$A_{1}^{n}$ associated with the $\Delta$ isobar. Both terms are 
important for the correct determination of  $A_{1}^{n}$. 
The term proportional to  $0.056$ decreases the absolute value 
of $A_{1}^{n}$ by about 6\%. 
Moreover, if $A_{1}^{^3{\rm He}}$ is negative, the second term 
proportional to $0.014$ would work in the same direction of 
decreasing of  $|A_{1}^{^3{\rm He}}|$. Since the term proportional to $0.014$ 
is always positive,  this means that  the true $A_{1}^{n}$ should turn 
positive at lower values of $x$ compared to the situation when the 
effect of the $\Delta$ is ignored. It is therefore vital to account for
these corrections in any extraction of $g_{1n}$ from $^3$He data in this
region \cite{Meziani}.

\section{CONCLUSION}
In conclusion we reiterate the importance of experiments aimed at
extracting information on the large-$x$ behaviour of the nucleon spin
structure -- especially that of the neutron. The results cast important
light onto the valence structure of the nucleon, testing our
understanding of SU(6) symmetry and the applicability of di-quark
models. In order to extract this information from nuclear targets
it is important to use a thorough theoretical analysis, including in
particular the effects of $\Delta$--isobars in the three-body wave
function.

\section*{ACKNOWLEDGEMENTS}
It is a pleasure to acknowledge the hospitality of the staff of ECT* and 
particularly S. Bass and W. Weise, during this very stimulating workshop.
I  would especially like to acknowledge my collaborators in the work
outlined in sections 3 and 4, F. Bissey, V. Guzey and M. Strikman. This work
was supported by the Australian Research Council and the University of
Adelaide.


%
\end{document}